\documentclass[epj,nopacs]{svjour}
\usepackage{graphics}
\usepackage{soul}
\usepackage{color}
\usepackage{amssymb}
\begin{document}
\title{Morphologies of compressed active epithelial monolayers}
\author{Jan Rozman\inst{1,2} \and Matej Krajnc\inst{1} \and Primo\v z Ziherl\inst{1,2}
}                     
%
%
\institute{Jo\v zef Stefan Institute, Jamova 39, SI-1000 Ljubljana, Slovenia \and Faculty of Mathematics and Physics, University of Ljubljana, Jadranska 19, SI-1000 Ljubljana, Slovenia}
\date{Received: date / Revised version: date}
%
\abstract{
Using a three-dimensional active vertex model, we numerically study the shapes of strained unsupported epithelial monolayers subject to active junctional noise due to stochastic binding and unbinding of myosin. We find that while uniaxial, biaxial, and isotropic in-plane compressive strains do lead to the formation of longitudinal, herringbone-pattern, and labyrinthine folds, respectively, the villus morphology characteristic of, e.g., the small intestine appears only if junctional tension fluctuations are strong enough to fluidize the tissue. Moreover, the fluidized epithelium features villi even in absence of compressive strain provided that the apico-basal differential tension is large enough. We analyze several details of the different epithelial forms including the role of strain rate and the modulation of tissue thickness across folds. Our results show that nontrivial morphologies can form even in unsupported, non-patterned epithelia. 
%
} 
\maketitle
%
\section{Introduction}
\label{sec:1}

The function of many animal organs strongly relies on the shape of the constituent epithelia. The foundations of the various organ forms are set in the early embryonic development, where the initially simple structures gain on complexity through different types of deformations including folding, wrinkling, in-plane stretching, and shear. Driven by a complex biochemical machinery, these processes are individually often discussed within the domain of materials science, but they are generally poorly understood in the biological context where they usually act together, and their synergistic and antagonistic effects remain theoretically largely unexplored. This is quite understandable as the parameter space increases with each additional process involved, which makes the analysis rather difficult.

Among the best-known epithelial morphologies are the intestinal villi. Villi form in embryos, e.g., upon a sequential differentiation of muscle tissues of the gut, which exert compressive stresses on the growing epithelium. These stre\-sses trigger buckling and villus formation through a few intermediate steps involving longitudinal and herringbone (zigzag) fold patterns~\cite{Shyer13}. So far, the mechanistic aspects of villus formation were interpreted in terms of several distinct processes such as tissue growth in confined space~\cite{Shyer13,Hannezo11,Ciarletta14} and mechanical apico-basal polarity~\cite{Storgel16}. Each of these processes may give rise to villus morphogenesis by itself although it is also possible that they act jointly so as to ensure robustness. However, the relative role and importance of the different possible mechanisms of villus formation remains an open question because most theoretical studies focus on a single one.

Many processes that contribute to epithelial morphogenesis can be theoretically investigated by describing the tissue as an elastic or viscoelastic continuum~\cite{Shyer13,Hannezo11,Ciarletta14,BenAmar13} but it is often advantageous to resort to cell-based computational schemes such as agent-based models~\cite{Galle05,Buske11}. The agent-based models build on cells whose state and response to signals from the environment are controlled by a set of rules and are thus well suited for the description of growth regulation, cell differentiation, spatial arrangement of the different cell populations, etc. At the same time, these schemes assume a very simple physical form of the cells, which are typically approximated by spheres, and they rely on a rather stripped-down representation of the mechanical cell-cell interaction. As a result, they can accommodate only a few non-specific morphogenetic modes such as folding due to rapid growth~\cite{Drasdo00,Drasdo05}. 

A geometrically and mechanically more accurate computational framework is provided by vertex models where each cell is represented by a polyhedron carrying a certain mechanical energy. Although still fairly coarse-grained, these models allow one to explore the contributions of the different structural elements of cells and tissues as well as to readily introduce various kinds of cell-level dynamics. While it is plausible that cell-resolution description may not be needed for many epithelial formations, the above features render the vertex models quite versatile and often advantageous over continuum theories because the energies and processes involved have a reasonably clear microscopic interpretation. In addition, the vertex models bypass the very choice of a continuum theory. The choice of such a theory poses a challenge since the deformations characteristic of epithelia are large, and hence even within the domain of elastic materials alone there exist a number of possibilities rather than a single one like in the small-deformation harmonic limit. At the same time, a continuum theory derived from a given vertex model in a bottom-up manner can be quite illuminating at a qualitative level as it may provide a connection between the various geometric quantities that describe the tissue shape~\cite{Krajnc15,Haas19,Rozman20}.

The main topological dynamic processes of interest in vertex models are neighbor-exchange T1 transition, cell division, and cell expulsion, which all result in in-plane rearrangements of cells and involve a certain kind of activity. Cell division generally takes place after the dividing cell has grown, often so as to double in volume, which evidently relies on metabolism, synthesis of biomolecules, etc. Irrespective of how the daughters arrange after the division, the local topology of the tissue is changed; the same applies to cell expulsion. As such, these two events contribute to tissue fluidization~\cite{Ranft10}. A third mechanism of fluidization is the T1 transition, where four cells around a given lateral side locally rearrange such that the two that initially shared the side no longer do so~\cite{Krajnc18}. This rearrangement generally requires climbing across an energy barrier~\cite{Bi14}, which is evidently an active process and can be implemented by a fluctuating tension representing binding and unbinding of junctional myosin~\cite{Curran17,Krajnc20}, by a more coarse-grained active T1 transition model~\cite{Krajnc18}, or in some other manner.

\begin{figure*}[t]
\includegraphics{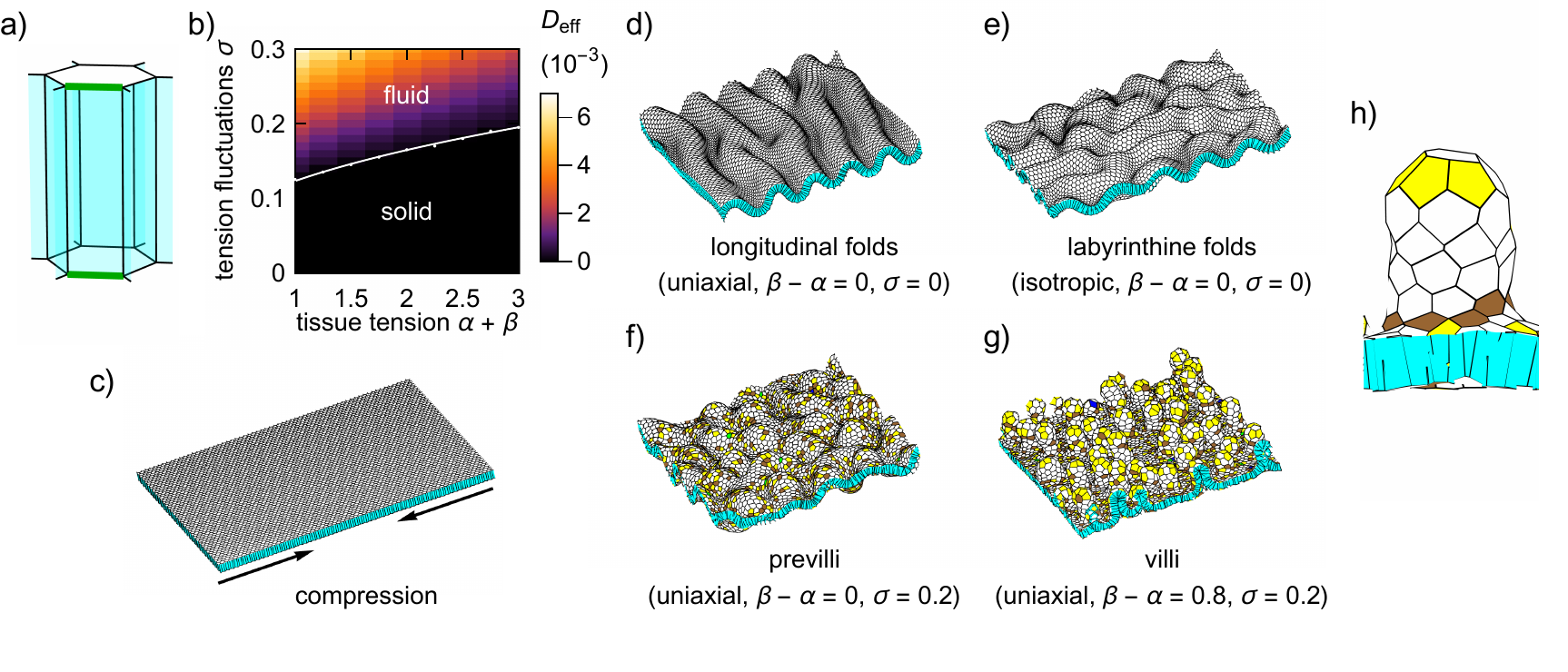}
\caption{Schematic of a six-coordinated cell in the model epithelial monolayer~(a). The fluctuating-tension edges at the apical (top) and the basal (bottom) side of the lateral surface (cyan) are highlighted by thick green lines. In panel~b we plot the diffusion coefficient of the cell as a function   of the tissue tension $\alpha+\beta$ and the magnitude of tension fluctuations $\sigma$. Superimposed on the color-coded plot is the numerically obtained location of the solid-fluid transition (white points) and the white line shows Eq.~(\ref{eq:9}). Panel~c shows an oblique view of the initial flat configuration, indicating the direction of uniaxial compressive strain. Also included are apical-surface views of representative $\alpha+\beta=2$ examples of the longitudinal~(d) and labyrinthine folds~(e) as well as the previlli~(f) and villi morphologies~(g); the strain mode, the differential tension $\beta-\alpha$, and the magnitude of tension fluctuations $\sigma$ are listed in each panel. Cells are colored according to the number of neighbors: Blue for 4, yellow for 5, white for 6, brown for 7, and green for 8. Lateral cell sides are shown in cyan. Panel~h shows a representative $\sigma=0.2$, $\alpha+\beta=2$, $\beta-\alpha={0.8}$ villus, with an excess of pentagonal (yellow) cells at the tip and an excess of heptagonal (brown) cells at the base.
}
\label{fig:1}
\end{figure*}

In a recent paper~\cite{Rozman20}, we used an active vertex model of epithelia to study the form of organoid-like epithelial shells, finding that the most elaborated, branched morphologies appear only if cell-level activity leading to in-plane cell rearrangements through T1 transitions is strong enough. The well-deve\-loped finger-like protrusions in the branched shapes are physically very reminiscent of the intestinal villi and crypts. As it appears possible that these protrusions also arise in a planar patch of an active epithelium provided that effect of the fixed-enclosed-volume constraint can be substituted by in-plane stress, we now explore the vertex model of epithelia under in-plane compressive strain. The main elements of the model as well as the representative morphologies obtained including a close-up of a villus are shown in Fig.~\ref{fig:1}. We find that while this strain does produce folded morphologies, villi or crypts appear only if two additional requirements are fulfilled: (i)~the tissue is completely fluidized and (ii)~the apico-basal differential tension exceeds a certain threshold. We also find that strain is not essential for the formation of the protrusions, which appear even in its absence as long as the above two requirements are met. Overall, while most previous studies relate the intestinal villi and crypts to the differential growth of the epithelium relative to  supporting tissues~\cite{Shyer13,Hannezo11,Ciarletta14}, our work demonstrates that they can form in an unsupported and non-patterned epithelial tissue.

The disposition of the paper is as follows: In Sec.~\ref{sec:2} we present the computational model used, elaborating all the details of the vertex model as well as our implementation of the tension-fluctuation scheme as the mechanism of active in-plane cell rearrangements. Section~\ref{sec:3} describes the key results whereas Sec.~\ref{sec:4}  discusses the state diagram and the relevance of our main findings in the context of experiments and existing theories. Section~\ref{sec:5} concludes the paper.    
\section{The model}
\label{sec:2}
We use a 3D vertex model where the tissue is represented by a monolayer sheet-like packing of $N_c$ polyhedral cells of identical fixed volumes $V_c$~\cite{Rozman20,Krajnc18,okuda15,misra16}. The apical and basal sides of the cells are polygons with the same number of vertices whereas the lateral sides are rectangular (Fig.~\ref{fig:1}). The shape of each cell is parametrized by the positions of vertices ${\rm \bf r}_j=(x_j,y_j,z_j)$, which obey the overdamped equation of motion
\begin{equation}
    \eta\dot{{\rm \bf r}}_j=-\nabla_j W+{\rm \bf F}_j^{(A)}\>.
    \label{eq:eqMotion}
\end{equation}
Here $\eta$ is an effective viscosity associated with the movement of vertices and 
\begin{equation}
    W = \sum_{i=1}^{N_{\rm c}} \left[\Gamma_a A_a^{(i)}+\Gamma_b A_b^{(i)}+\frac{\Gamma_l}{2}A_l^{(i)}\right]\>
    \label{eq:energy}
\end{equation} 
is the energy of the epithelial sheet;  $A_a^{(i)},A_b^{(i)}$, and $A_l^{(i)}$ are the areas of the apical, basal, and lateral sides of the $i$-th cell and $\Gamma_a,\Gamma_b,$ and $\Gamma_l$ are the corresponding surface tensions, each of them assumed to be identical in all cells; the factor of 1/2 in the third term arises because each lateral side is shared between two cells.

The second term on the right-hand side of Eq.~(\ref{eq:eqMotion}), ${\bf F}_j^{(A)}$, is the active force describing the noise at cell-cell junctions which arises from the stochastic turnover dynamics in the junctional myosin. These dynamics take place at the lateral cell-cell contacts, close to the apical cell surface~\cite{Lecuit07}. Since the lateral sides of our model cells are not discretized along the apico-basal axis and T1 transitions always occur simultaneously on the apical and the basal cell side, we associate these fluctuations with the apical and the basal edge of cell-cell junctions such that 
\begin{equation}
    {\rm \bf F}_j^{(A)}=-\sum_{k=1}^{N_e}\gamma_k(t)\nabla_j\left[ L_a^{(k)} + L_b^{(k)}\right]\>.
    \label{eq:activeForces}
\end{equation}
Here $L_a^{(k)}$ and $L_b^{(k)}$ are apical and basal edge lengths of the $k$-th cell-cell junction, respectively, and the sum goes over all $N_e$ junctions. Junctional line tensions $\gamma_k(t)$ are described by the Ornstein-Uhlenbeck process
\begin{equation}
    \frac{{\rm d}\gamma_k(t)}{{\rm d}t}= -\frac{1}{\tau_m}\gamma_k(t) +\xi_k(t)\>,
\end{equation}
where $1/\tau_m$ is the turnover rate of molecular motors and $\xi_k(t)$ is the Gaussian white noise with properties $\left <\xi_k(t)\right >=0$ and $\left <\xi_k(t)\xi_{k'}(t')\right>=(2\sigma^2/\tau_m)\delta_{k,k'}\delta(t-t')$, $\sigma^2$ being the long-time variance of tension fluctuations~\cite{Rozman20,Curran17,Krajnc20}.

As a mechanism of tissue fluidization, junctional tension fluctuations are one of the key elements of our model: If the  magnitude of fluctuations exceeds a critical value, the monolayer undergoes a transition from solid-like to fluid-like state~\cite{Krajnc20}. In the latter, cells can move within the monolayer and this motion has a dramatic and often far-reaching effect on the mechanical features as well as on the shape of the tissue. For example, it can trigger large-scale deformations and cell rearrangements during embryonic development~\cite{Curran17,Mongera18} but it can also be an essential factor in the formation of the more elaborated 3D tissue morphologies in highly constrained systems such as organoids~\cite{Rozman20}. 
\subsection{Surface tensions}
We first non-dimensionalize the model. We use the la\-teral tension $\Gamma_l$ as the unit of tension whereas the natural choice of the unit of surface area is based on cell volume $V_c$ and reads $V_c^{2/3}$. The dimensionless energy is given by 
\begin{equation}
    w = \sum_{i=1}^{N_c} \left[\alpha a_a^{(i)}+\beta a_b^{(i)}+\frac{1}{2}a_l^{(i)}\right],\label{eq:dimensionless}
\label{eq:w}    
\end{equation}
where 
\begin{equation}
    \alpha=\frac{\Gamma_a}{\Gamma_l}
\end{equation}
and
\begin{equation}
   \beta=\frac{\Gamma_b}{\Gamma_l}
\end{equation}
are the reduced apical and basal surface tensions, respectively, whereas $a_a^{(i)}=A_a^{(i)}/V_c^{2/3}$ is the reduced area of the apical side (and analogously for $a_b^{(i)}$ and $a_l^{(i)}$). The reduced apical and basal tensions jointly determine the preferred shape of individual cells. In particular, the tissue tension $\alpha+\beta$ controls the cell width-to-height ratio: For $\alpha+\beta$ considerably smaller than 1 this ratio is large which corresponds to squamous tissues whereas for $\alpha+\beta$ much larger than 1 it is small which represents columnar tissues. On the other hand, the apico-basal differential tension $\beta-\alpha$ favors cells of conical shape which may lead to buckling at the tissue scale. Unless stated otherwise, we report results for columnar model tissues~(Fig.~\ref{fig:1}a) with $\alpha+\beta=2$, giving the cell width-to-height ratio of about 0.5, and we vary the differential tension $\beta-\alpha$ so as to explore the contribution of the apico-basal polarity to the formation of various morphologies. Without loss of generality, we restrict the discussion to $\beta>\alpha$ so that the differential tension $\beta-\alpha$ is positive. The obtained non-trivial morphologies that feature basal constrictions and  protrusions penetrating into the lumen can be translated into those characterized by apical constrictions and recesses simply by swapping the sign of the differential tension. For example, the villus morphology at $\beta-\alpha=0.8$ in Fig.~\ref{fig:1}g also represents the crypt morphology at $\beta-\alpha=-0.8$, with Fig.~\ref{fig:1}g showing the basal rather than the apical view of the latter.
\subsection{Compressive strain}
We investigate tissues consisting of $N_c$ cells with periodic boundary conditions along the two in-plane directions denoted by $x$ and $y$. The in-plane compression is simulated by decreasing the linear dimensions of the simulation box $L_x$ and $L_y$ either along a single in-plane axis so as to ge\-nerate uniaxial strain or along both of them by the same amount, which gives rise to isotropic strain. In our model, these dimensions are decreased linearly in time so that 
\begin{equation}
L_\mu(t)=L_\mu^{(0)}\left(1-\varepsilon_\mu\frac{t}{T}\right),
\end{equation}
where $L_\mu^{(0)}$ is the initial dimension of the box in question, $\mu$ is either $x$ or $y$, $T$ is the total simulation time, and $\varepsilon_\mu$ is the final in-plane strain along the corresponding direction. Of course, $\varepsilon_\mu$ corresponds to the physical definition of strain only if no out-of-plane buckling occurs, which is not the case in our simulated morphologies.

\subsection{Tension fluctuations}
The competition between cell surface tension and friction with the environment determines the typical time scale of vertex movements given by $\tau_0=\eta/\Gamma_l$, which is here used as the unit of time. After introducing the dimensionless time $t/\tau_0\to t$, the equation of motion for the dimensionless vertex positions ${\rm \bf r}_j/V_c^{1/3}\to{\rm \bf r}_j$ reads $\dot{{\rm \bf r}}_j=-\nabla_j w+{\rm \bf f}_j^{(A)}$, where ${\rm \bf f}_j^{(A)}$ is the dimensionless active force on the $j$-th vertex measured in units of $\Gamma_lV_c^{1/3}$. In turn, the dynamical equation for dimensionless junctional tensions $\gamma_i/\left (\Gamma_l V_c^{1/3}\right )\to\gamma_i$ reads $\dot{{\gamma}}_i(t)=-{\gamma}_i(t)/\tau_m+\xi_i(t)$, where the long-time variance of dimensionless line tension fluctuations $\xi_i$ is given by $\sigma^2/\left (\Gamma_l^2 V_c^{2/3}\tau_0^{-2}\right )\to\sigma^2$ and the dimensionless inverse myosin turnover rate is $\tau_0/\tau_m\to1/\tau_m$. Here we study the effect of the magnitude of junctional tension fluctuations $\sigma$ but we keep the myosin turnover rate fixed at $1/\tau_m=1$. 

To determine the critical magnitude of junctional tension fluctuations $\sigma^*$ where the solid-fluid transition takes place as previously discussed and shown to hold in a 2D model~\cite{Krajnc20}, we simulate fluctuations in a flat non-stressed epithelial patch in absence of differential tension, i.e., at $\alpha = \beta$. During these simulations, we affix the basal sides of the cells to a flat plane so as to prevent buckling out of the plane. Starting from a honeycomb arrangement, we allow the simulation to run at a given $\sigma$ from $t=-10^3$ to $t=0$. Following Ref.~\cite{Krajnc20}, we then continue to run the simulation to $t = 10^3$ and we measure the mean square displacement of cell centers given by
\begin{equation}
    \mathrm{ MSD}(t)=(1/N_c)\sum_{i=1}^{N_c}\left|{\bf{r}}_b^{(i)}(t) - {\bf r}_b^{(i)}(0) \right|^2, 
\end{equation}
where ${\bf r}_b^{(i)}(t)$ is the in-plane position of the center of basal side of the $i$-th cell. Using MSD as a measure of cell movement, we can then determine the effective diffusion coefficient $D_{\rm eff}={\rm MSD}(t)/(4t)|_{t\rightarrow \infty}$. Similar to what was observed in the 2D version of the vertex model~\cite{Krajnc20}, the diffusion coefficient only takes finite values above a certain threshold of tension fluctuations. At the tissue tension of $\alpha+\beta=2$ which pertains to most results presented in this paper, the threshold is $\sigma=\sigma^*\approx 0.165$. For $\alpha+\beta$ between 1 and 3, the critical $\sigma^*$ increases with $\alpha+\beta$ as
\begin{equation}
    \sigma^*(\alpha+\beta) \approx 0.123\left(\alpha+\beta\right)^{0.42}.
   \label{eq:9}
\end{equation}

\subsection{Implementation}
Within the computational scheme outlined above, we explore the parameter space of our model by performing a series of simulation runs; rather than embarking on a systematic scan of this space, we focus on the most interesting effects at work. In each run, we start with a flat monolayer consisting of $N_c$ identical cells that form a hexa\-gonal prismatic honeycomb, and we simulate its evolution upon compression until the final strain is reached at a time $T$. Here we examine a honeycomb that consists of 64~rows of 64~cells arranged in a rectangular patch with an aspect ratio of $1 : \sqrt{3}/2$ equal to the ratio of the long and short diagonal of a regular hexagon; the total number of cells is $N_c=4096$. The compression time measured in units of $\tau_0$ is $T=1000$ unless stated otherwise, and $\alpha+\beta=2$, again unless stated otherwise. In the isotropic compression mode $\varepsilon_x=\varepsilon_y=15~\%$, whereas in the uniaxial compression mode $\varepsilon_x=27.75~\%$ and $\varepsilon_y=0$ so that the projected areas of the isotropically and the uniaxially strained tissues given by $L_x(T)L_y(T)$ are identical. To facilitate the evolution of the tissue from the very regular initial honeycomb state, we break its apico-basal symmetry by displacing each vertex at $t=0$ in a random direction such that the displacements are normally distributed around zero with a standard deviation of $0.03V_c^{1/3}$.

An essential element of our scheme is the implementation of the T1 transition, which is executed in two steps as described in the following. If the length of a junction (defined as the average of the corresponding apical and basal edge length) falls under a threshold value $l_0 = 0.01$ and has decreased in length since the previous time-step, the junction is first merged into a fourfold vertex, giving rise to a local 4-cell rosette arrangement. The rosette is then maintained for a short time interval of 0.02; this is the second step. After this time, it is resolved into two three-way vertices, displacing each from the pre-resolution position by 0.0005. After the T1 transition, a new line tension is assigned to the junction from a normal distribution with mean 0 and standard deviation $\sigma$.

Lastly, we note that our scheme does not include the steric repulsion between the cells; it is therefore possible for non-neigboring cells to overlap. In practice, this issue may arise only on the basal (inner) side of the villus morphologies, and this may only happen after the villi are already established. As such, self-overlap may slightly affect the detailed shape of the villi but not their existence or the qualitative conclusions of the paper.
\section{Four morphologies}
\label{sec:3}
Our scheme produces four types of morphologies of compressed epithelial monolayers: The longitudinal and the labytrinthine folds depend on the type of compression (uniaxial and isotropic, respectively) 
whereas previlli and villi are independent of the details of compression but require that the tissue is fluidized by junctional fluctuations. 
%
\subsection{Longitudinal folds}
\label{sec:3a}

\begin{figure}[t!]
\centering
 \includegraphics{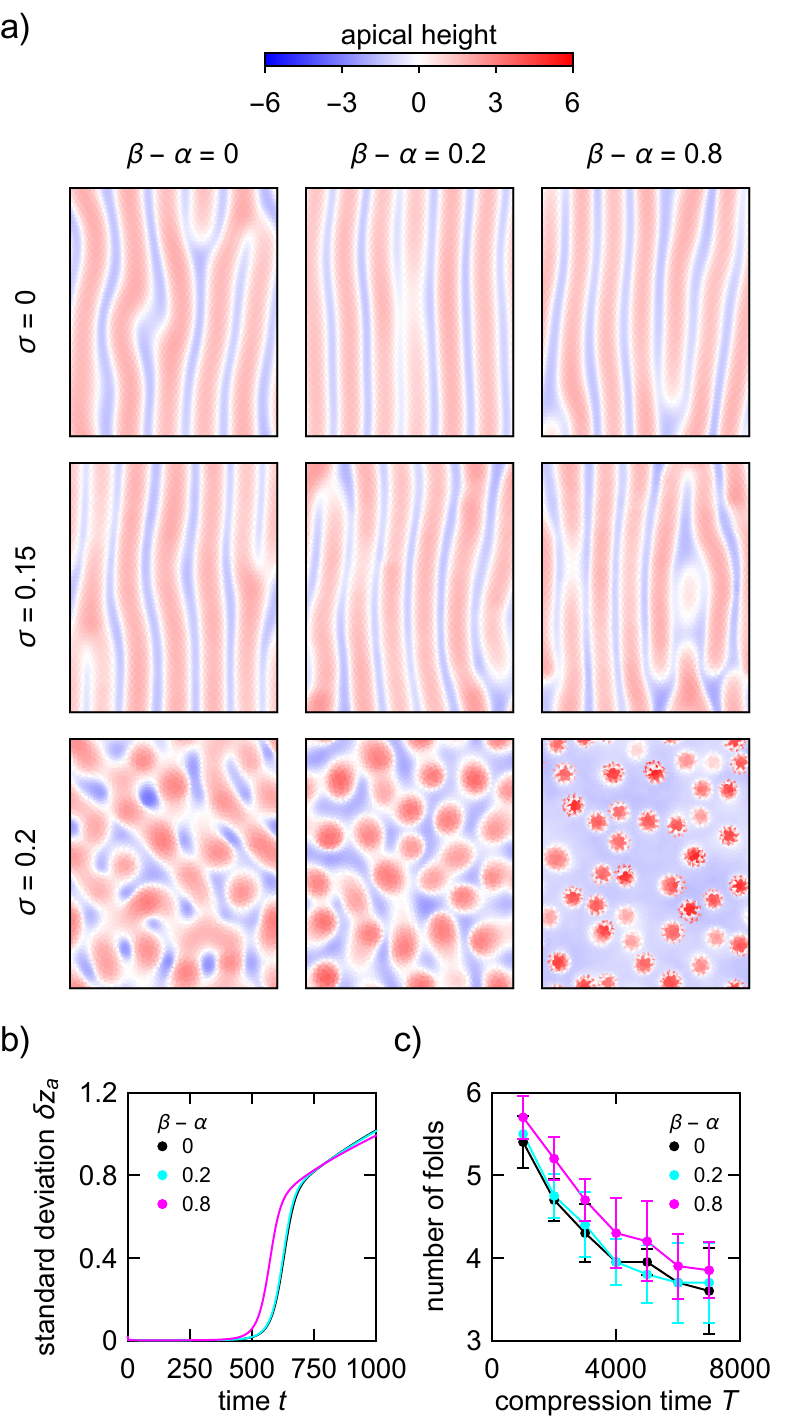} 
\caption{Height profiles of the apical surface of epithelial morphologies formed by uniaxial compression along the $x$ axis by 27.75~\% at $\sigma=0, 0.15,$ and $0.2$ and $\beta-\alpha=0, 0.2,$ and $0.8$~(a). The top and middle row show longitudinal folds seen in the solid-like tissue whereas the bottom row contains the protovilli ($\beta-\alpha=0$ and 0.2) and the villi ($\beta-\alpha=0.8$) morphologies. The profiles show the positions of the centers of apical sides, with 0 corresponding to the average height. In panel~b we plot the temporal profile of the standard deviation of the height of apical vertices in three $\sigma=0$ solid-like tissues, averaged over ten simulation runs, which indicates that buckling appears at a time of $t\approx500$. Panel~c shows the number of folds in a $\sigma=0$ solid-like tissue vs.~compression time $T$ at three different~$\beta-\alpha$, averaged over ten simulation runs; error bars indicate standard deviation. Note that the precise count of folds is somewhat subjective, as some individual folds are branched or do not stretch across all of the patch.
}
\label{fig:2}
\end{figure}

The simplest nontrivial epithelial shape occurs in a uniaxially compressed tissue with a level of junctional tension fluctuations below the solid-fluid transition. Much like a rod compressed in the lengthwise direction, the solid-like epithelium undergoes an Euler instability and a buckling transition, assuming a corrugated shape that features a series of parallel folds running perpendicular to strain direction (Fig.~\ref{fig:1}d, Fig.~\ref{fig:2}a, and Movie~S1). In our scheme, buckling does not develop immediately after the strain is applied but only after a certain finite time. The development of the deformation can be appreciated by monitoring the deviation of the apical vertices from their average height $\delta z_a$, which is close to 0 in a flat tissue before buckling and finite in a modulated tissue. Figure~\ref{fig:2}b shows that at the magnitude of the initial random displacement of the vertices used here, the deformation of a $\sigma=0$ solid-like tissue rapidly forms around time $t\approx500$ as witnessed by the step-like profile of $\delta z_a(t)$; this buckling time is slightly shorter at larger differential tensions $\beta-\alpha$. After that, the deformation continues to grow which shows that the response of the tissue to strain is dynamic. As may be anticipated, the dynamics involves fold coarsening; here we collect and report tissue shape right after compression time $T=1000$. Notably, the time at which buckling occurs does significantly depend on the scale of the initial perturbation of the vertices at time $t = 0$ (Fig.~{S1}).

The $\sigma=0$ fold morphologies obtained at a compression time of $T=1000$ shown in the top row of Fig.~\ref{fig:2}a do not seem to be very sensitive to the apico-basal differential tension $\beta-\alpha$. However, the number of folds decreases with the compression time $T$ as shown in Fig.~\ref{fig:2}c. The effect of the strain rate is qualitatively similar to folding in a toy model of a fixed-end growing tissue consisting of circular cells where the number of folds decreases with the duration of the cell cycle; that is, in a slowly dividing tissue the number of folds is smaller than in a rapidly dividing one~\cite{Drasdo00}. The data shown in Fig.~\ref{fig:2}c also illustrates the subdominant dependence of the number of folds on the differential tension. Despite the large error bars, our data suggest that the number of folds slightly increases with $\beta-\alpha$. This effect is not surprising since the differential tension promotes the constriction of the basal cell sides, inducing a preferred truncated-pyramid cell shape which is more easily accommodated in a folded than in a flat tissue~\cite{Storgel16,Krajnc13}. Lastly, we note that at a finite yet subcritical magnitude of junctional tension fluctuations the number of folds in a uniaxially compressed epithelial monolayer is more or less the same as at $\sigma=0$ as illustrated by the $\sigma=0.15$ sequence shown in the middle row of Fig.~\ref{fig:2}a. 

\begin{figure}[t]
\centering
\includegraphics{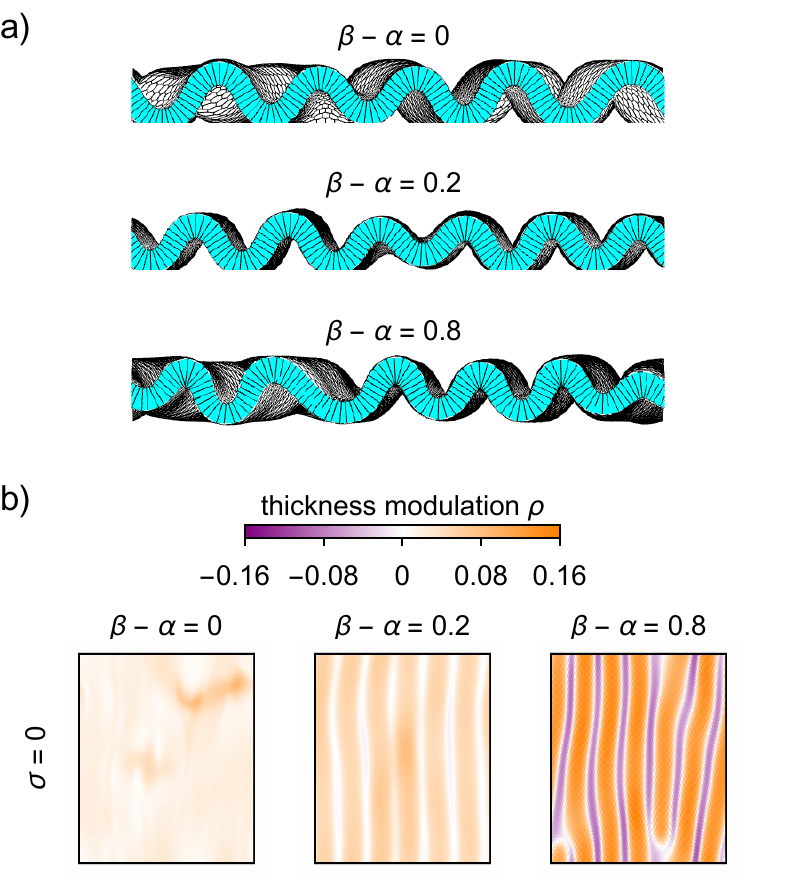} 
\caption{Side views of the $\sigma=0$ longitudinal folds in a uniaxially strained model tissue at apico-basal differential tensions $\beta-\alpha=0,0.2,$ and 0.8~(a). Thickness modulation in these three fold morphologies~(b). }
\label{fig:6}
\end{figure}

The longitudinal folds are also interesting because they allow us to explore the thickness-curvature coupling~\cite{Storgel16,Rozman20} as transparently as possible. The effect of the coupling can be quantified by thickness modulation
\begin{equation}
\rho = \frac{h - h_0}{h_0},    
\end{equation}
where $h$ is the height of a cell defined as the distance between the centroids of its apical and basal sides whereas $h_0 = \left(\sqrt{3}/2\right)^{-1/3}\left(\alpha + \beta\right)^{2/3}$ is the equilibrium cell height in a flat regular honeycomb tissue~(Fig.~\ref{fig:6}b)~\cite{Krajnc18}.

The mechanism behind thickness modulation goes as follows. In a tissue with a given apico-basal differential tension $\beta-\alpha>0$, the preferred minimal-energy shape of cells is a truncated pyramid with the basal side smaller than the apical side because of a larger basal tension; such a shape has a certain curvature. A fold includes cells of preferred shape but it must also consist of those with the opposite curvature because the integrated curvature of its contour must vanish. In the latter, the basal side is larger than the apical side, which is energetically unfavorable. The local energy increase can be minimized by moving the apical and the basal side closer to each other (accompanied by the expansion of the in-plane diameter due to the fixed-volume constraint). In the elastic theory, this phenomenon shows as the thickness-curvature coupling term with a modulus proportional to  $\beta-\alpha$~\cite{Krajnc15,Rozman20}.  

In Fig.~\ref{fig:6}a we compare three examples of fold morphologies at $\beta-\alpha=0,0.2,$ and 0.8. One can spot the constrictions on the basal (bottom) sides of crest cells and the more broad and open shape of the grooves present in the $\beta-\alpha=0.8$ folds and absent in the symmetric $\beta-\alpha=0$ folds. In the $\beta-\alpha=0.8$ folds, the thickness modulation along the waveform---thin tissue in the grooves, thick tissue in the crests---can be easily seen. A detailed inspection shows that the magnitude of the modulation in these fold morphologies shown in Fig.~\ref{fig:6}b ranges from $\approx 10~\%$ of the equilibrium height in a flat honeycomb at $\beta-\alpha=0$ to $\approx 25~\%$ at $\beta-\alpha=0.8$, which emphasizes that the effect is proportional to $\beta-\alpha$. 

\subsection{Previlli}
\label{sec:3c}
In epithelia that are completely fluidized by junctional tension fluctuations, buckling caused by strain takes on a different from than in solid-like tissues even at small apico-basal differential tension (Fig.~\ref{fig:1}f, Fig.~\ref{fig:2}a, and Movie~S2). Instead of the relatively evenly spaced and regular folds, the uniaxially strained fluidized epithelium forms a structure consisting primarily of bumps and pits, i.e., protrusions pointing up and down. In addition to these spatially isolated features, the nearby bumps and pits fuse into uneven short crests and grooves, respectively, essentially doublets or triplets of individual protrusions as illustrated by the $\sigma=0.2,\beta-\alpha=0$ morphology in the bottom row of Fig.~\ref{fig:2}a. Interestingly, there seems to be no preferred ave\-rage in-plane orientation of these features, which suggests that this morphology does not depends on the direction of the applied strain.  

The bumps and pits are similar to villi and crypts, respectively, in that they are local rather than extended spatial features of the epithelium, but they are much less developed.  At the same time, as the apico-basal differential tension is increased, the bumps become more sharply defined to the point where they can be viewed as villi, and the space between them is flattened out. As such the bumps-and-pits morphology may be referred to as the previlli (or, equivalently, precrypts if $\beta-\alpha$ is negative).     

\subsection{Villi}
\label{sec:3d}
The third epithelial morphology are the villi observed in completely fluidized tissues at a large enough apico-basal differential tension (Fig.~\ref{fig:1}g and Movie~S3). As shown by the bottom row of Fig.~\ref{fig:2}a, at $\beta-\alpha=0.2$ villus-like bumps protrude from most of the model tissue but some fold-like formations are still present, whereas at $\beta-\alpha=0.8$ the villi are completely developed and so is the intervening flat intervillous table. The spatial distribution of the villi appears to be random and unrelated to the direction of strain, which is an indication that the roles of strain and activity in villus formation are quite disjunct. In the close-up of a villus in Fig.~\ref{fig:1}h, we also observe the non-trivial in-plane topological structure with a excess of seven-coordinated cells in the base and an excess of five-coordinated cells at the tip. This characteristic distribution has been first reported in the branches of organoid-like epithelial shells~\cite{Rozman20} but is apparently an intrinsic feature of protrusion-forming vertex models.

The results presented here do not depend very much on the tissue tension $\alpha + \beta$ in a significant range around $\alpha + \beta=2$. The three discussed morphologies--- longitudinal folds, previlli, and villi---also arise at significantly lower ($\alpha+\beta = 1$) as well as higher ($\alpha+\beta=3$) tissue tensions under an appropriate rescaling of the differential tension $\beta - \alpha$ and tension fluctuation $\sigma$ (Fig.~{S2}). The same is also true of the labyrinthine morphology introduced in Sec.~\ref{sec:3b}. Furthermore, our choice to compress the tissue along the $x$ axis does not significantly affect the observed morphologies; the results are essentially the same for model tissues under compression along the $y$ axis (Fig.~{S3}).

Our model produces an inconsequential artifact in the villus morphologies where the basal sides of many cells at the tips are very small or even vanish. In these pyramidal cells, one or more basal vertices can move towards the apical side so as to increase the length of the basal edges at unchanged small or zero area of the basal side. This may be energetically favorable because of the fluctuating, occasionally negative basal edge tension (Fig.~{S4}a). As a result, several spikes may appear on the basal side of the villi tips (Fig.~{S4}b). 

The basal spikes may be suppressed by including an energy penalty on the very small basal areas. This auxi\-liary term reads $\sum_i k_b\left[A_b^{(i)}-A_0\right]^2{\mathrm H}\bigl(A_0 - A_b^{(i)}\bigr)$, where the sum runs over all cells, $k_b$ is the modulus, $A_b^{(i)}$ is the basal area of the $i$-th cell, $A_0$ is the threshold area under which the penalty is activated, and H is the Heaviside function~\cite{Sui18}; here $k_b = 200$ and $A_0$ is set to 30~\% of the equilibrium basal area in a flat regular honeycomb tissue. With this modification, the villi are almost completely free of the spikes on the basal side (Fig.~{S4}c).

\subsection{Labyrinthine folds}
\label{sec:3b}
\begin{figure}[b]
\centering
 \includegraphics{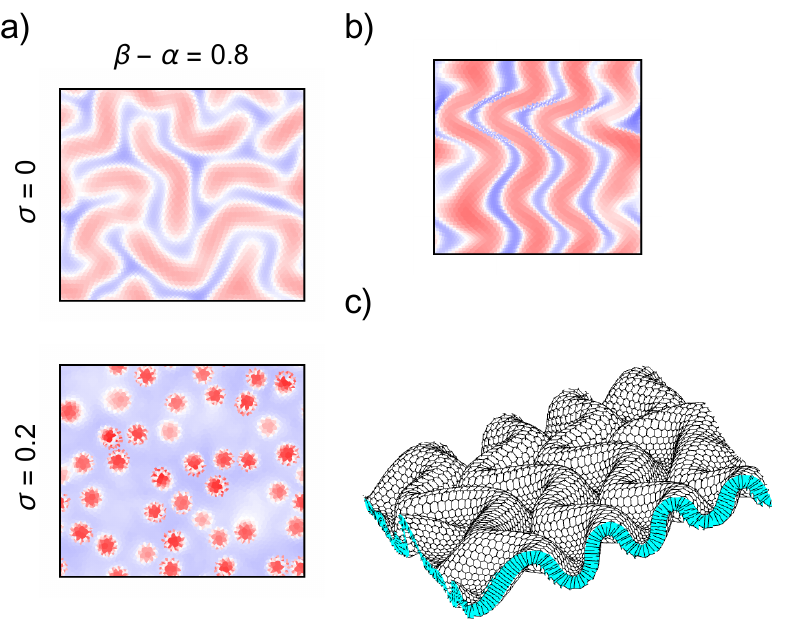} 
\caption{Height profiles of the apical surface of an isotropically compressed epithelial monolayer with a strain of 15~\% and a magnitude of junctional tension fluctuations $\sigma=0$ and 0.2~(a), displaying the labyrinthine and the villus morphologies, respectively. The color legend is the same as in Fig.~\ref{fig:2}a. Panel~b shows the $\sigma=0,\beta-\alpha={0.2}$ herringbone folds obtained by sequential biaxial compression starting from the configuration in Fig.~\ref{fig:2}a and then applying strain along the $y$ axis by {22~\%} at a compression time of $T'=2200$; panel~c is an oblique 3D view of the herringbone pattern.
}
\label{fig:4}
\end{figure}

After analyzing uniaxially compressed tissues, we also consider two other strain modes. Under isotropic compression with  $\varepsilon_x=\varepsilon_y=15~\%$, a solid-like $\sigma<\sigma^*$ monolayer forms a disordered labyrinthine arrangement of folds (Fig.~\ref{fig:4}a, Fig.~{S5}, and Movie~S4). Here the folds do not span across the whole patch but are of finite length, usually bent rather than straight, and lack a preferred orientation. The detailed shape of the folds depends only slightly on the apico-basal differential tension.
An even more interesting result of the isotropic-strain case is the $\sigma=0.2,\beta-\alpha=0.8$ villus morpho\-logy shown in the bottom row of Fig.~\ref{fig:4}a, which is essentially the same as that obtained under uniaxial strain (Fig.~\ref{fig:2}a). This finding confirms our speculation from Sec.~\ref{sec:3d} that villus formation does not depend critically on the applied compressive strain but that it is conditioned by cell-level activity and differential tension alone.

Finally, we also studied sequential compression of the mo\-del tissue monolayer in two perpendicular directions to see if our model can also reproduce the common herringbone folds~\cite{Shyer13,Hannezo11}. Given that the villus morphology does not depend on the direction of strain, we focus on the solid-like tissue with $\sigma=0$. For clarity, we use the $\varepsilon_x=27.75~\%$, $T=1000$ fold state with $\beta-\alpha=0.2$ from Fig.~\ref{sec:2}a as the first stage, intentionally choosing a starting configuration with well-developed longitudinal folds. We then apply strain along the existing folds, reaching a final value of {$\varepsilon_y=22~\%$} in time $T'=2200$. We find that the sequential compression as described here indeed leads to a herringbone pattern of zigzagging folds shown in Figs.~\ref{fig:4}b and~c.

\section{Discussion}
\label{sec:4} 

\begin{figure}[t]
\centering
 \includegraphics{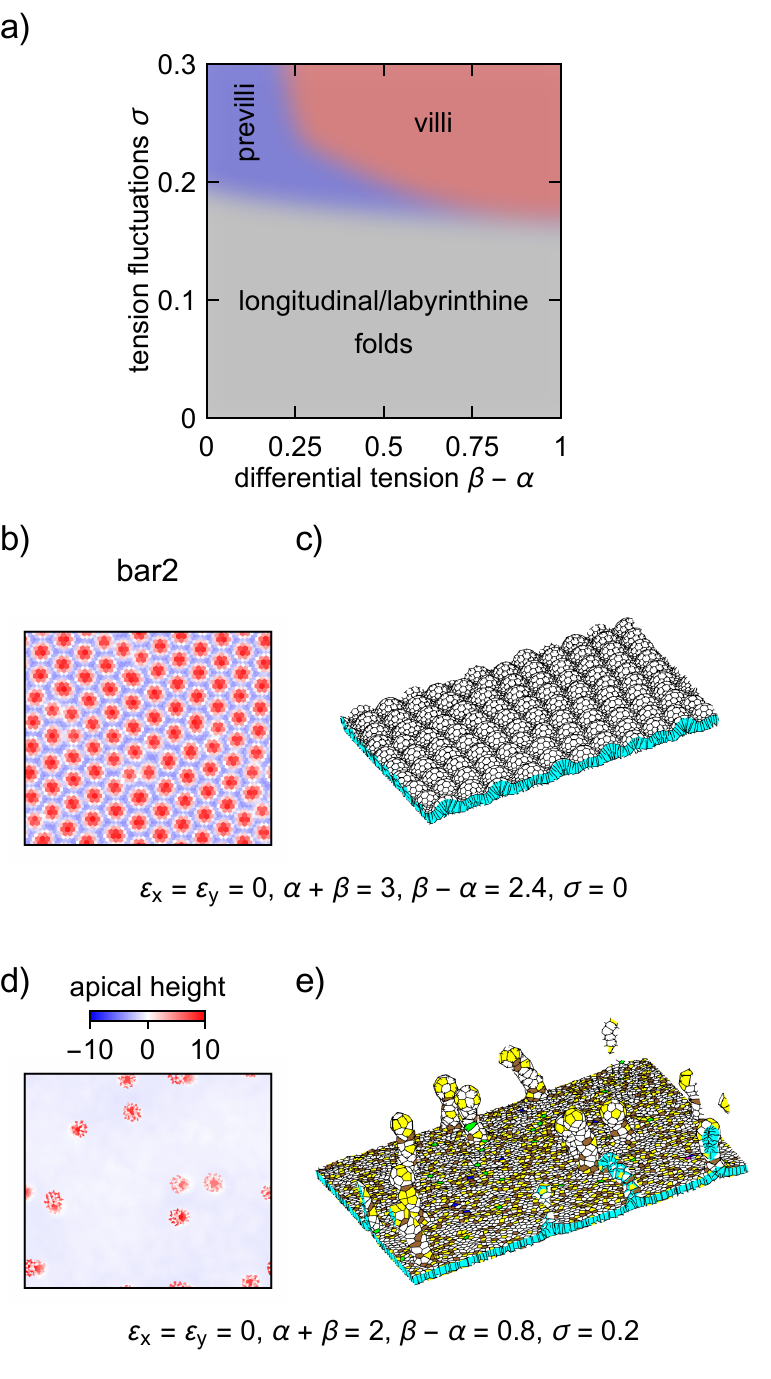} 
\caption{State diagram of a compressed epithelial monolayer with $\alpha+\beta=2$ at isotropic strain with $\varepsilon_x=\varepsilon_y=15~\%$ combined with that at equivalent uniaxial strain with $\varepsilon_x=27.75~\%$ and $\varepsilon_y=0$~(a).  Panels~b and c show an array of previllous bumps in a solid-like tissue with a large differential tension at $\varepsilon_x=\varepsilon_y=0$ with $\sigma=0$ at time $t=10000$, whereas in panels~d and e we plot an example of the villus morphology induced by tension fluctuations in an unstrained $\varepsilon_x=\varepsilon_y=0$ flat tissue at $t=3000$.
}
\label{fig:5}
\end{figure}

The different morphologies can be arranged in a  state diagram shown in Fig.~\ref{fig:5}a, where we combine the results pertaining to the different types of strains so as to better emphasize the individual roles of strain, apico-basal differential tension, and activity. This allows us to group the two low-activity fold morphologies---longitudinal and labyrinthine---in a single domain extending to $\sigma\lesssim\sigma^*\approx{0.17}$ (the threshold $\sigma$ is slightly higher at lower $\beta-\alpha$, but this effect is not very pronounced). On the other hand, the $\sigma>\sigma^*$ domain is occupied by previlli and by villi at small and large differential tensions, respectively. In the explored region of the phase space, villi appear at $\beta-\alpha\approx 1$ just above the solid-fluid transition. As $\sigma$ is increased, the boundary shifts rapidly to significantly smaller differential tensions down to $\beta-\alpha\approx0.4$; from this point on, the boundary is relatively vertical, reaching $\beta-\alpha\approx0.2$ by $\sigma=0.3$, the largest magnitude of junctional tension fluctuations considered. We note that the distinction between the previlli and the fully developed villi is quite arbitrary, hence the blurred boundary in the state diagram. More specifically, as the differential tension increases, the symmetry between the previllous bumps and pits breaks so that the former exceed the latter. Afterwards, the bumps become more clearly defined, producing the villi (Fig.~{S6}). Similarly, the boundary between the solid-like folds and the previlli can also be somewhat unclear, because, e.g., some of the longitudinal orientation is retained at $\sigma$ slightly above 0.165.

The state diagram in Fig.~\ref{fig:5}a  emphasizes that villus formation relies on a strong enough junctional tension fluctuations as well as on a large enough apico-basal differential tension. The diagram corresponds to isotropic in-plane strain with $\varepsilon_x=\varepsilon_y=15~\%$, and to the equivalent uniaxial strain with $\varepsilon_x=27.75~\%$ and $\varepsilon_y=0$, but we note that it remains qualitatively the same at any moderate compression. 

The natural question to ask is to what extent are these results affected by compression. To examine this effect, we scanned the parameter space so as to find that an initially flat $\sigma=0$ model tissue placed in a perfectly fitting box with $\varepsilon_x=\varepsilon_y=0$ did undergo a deformation after a large enough differential tension was turned on, forming a previllous array; an example with $\alpha+\beta=3$ and $\beta-\alpha=2.4$ is shown in Figs.~\ref{fig:5}b and~c. Upon deformation, the area of the tissue is evidently increased so that the fixed box size conceivably imposes a dilatational rather than a compressive strain. While we did not investigate how this morphology changes if the box size is varied, the existence of previlli at $\varepsilon_x=\varepsilon_y=0$ demonstrates that this morphology may be induced by differential tension alone; note that when studying the effects of compression, we only focused on moderate differential tensions up to $\beta-\alpha=1$, which are insufficient to produce previllous bumps at $\alpha+\beta=2$ in absence of strain. Likewise, compression is  not essential for the existence of villi. This is illustrated by Figs.~\ref{fig:5}d and e showing villi that developed at $\varepsilon_x=\varepsilon_y=0$ from a fluid-like flat tissue with $\sigma=0.2$, $\alpha+\beta=2,$ and $\beta-\alpha=0.8$, which is well within the range covered by the state diagram in Fig.~\ref{fig:5}a. The villi are present albeit in smaller numbers than in the compressed tissue with identical material parameters in Fig.~\ref{fig:1}g. These results lead us to conclude that within our model, previllous morphologies may arise in fludized epithelia by themselves or due to differential tension in solid-like tissues (where their formation could conceivably be aided by strain) whereas villi appear only in fluidized tissues at a large enough differential tension.

Many of our findings agree well with experimental studies of epithelial morphogenesis and they complement certain aspects of existing theories. In one of the key studies in the field~\cite{Shyer13}, the step-by-step development of the villi covering the gut tube of the chick embryo was associated to the differentiation of the distinct muscle layers of the gut. The initially smooth luminal epithelium forms longitudinal folds after the muscles that compress it circumferentially have developed, and the lengthwise muscles that appear next then deform the folds into a herringbone pattern and further into an array of previllous bumps. This sequential morphogenetic transformation was ni\-cely reproduced by a tailor-made composite structure where the muscle was represented by a rigid constraint whereas the epithelium as well as the mesenchymal layer between the muscle and the epithelium were treated as a suitable hyperelastic material. In this model, the epithelium is mostly seen as a passive element which conforms to the buckled shape of the mesenchyme resulting from the compressive strain due to the muscle around it. 

Our results suggest that a very similar progression of the morphologies can conceivably be obtained in a sub\-strate-free epithelium. We show that the first two steps of the process---the formation of longitudinal folds and zigzags---can be obtained by a suitable sequential biaxial strain in the solid-like regime. The third step, the formation of villi from zigzags, was not explicitly tested in out model, but based on the above findings it seems possible that a suitable combination of tissue fluidization and an increase in differential tension would produce such an outcome. In a similar manner, our results also relate to the buckled morphologies generated by differential isotropic~\cite{Ciarletta14} and anisotropic~\cite{BenAmar13} growth of an epithelium on a supporting tissue as such growth produces strain.

More importantly, our model also provides an interpretation for the development of the mouse-gut villi which form directly from the smooth epithelium. In Ref.~\cite{Shyer13}, this particular morphology was attributed to a softer endoderm compared to that in the chick embryo, but our results suggest that the mouse-gut-like villi may also result from a strong enough activity, a suitable compressive strain, and a large enough differential tension. We also note that the mouse-gut villi do not form an ordered array but are randomly distributed across the epithelium. This agrees with our villus morphologies as illustrated by bottom-right profile in Fig.~\ref{fig:2}a but not with the regular patterns obtained in models that build exclusively on elastic instability~\cite{Shyer13,Ciarletta14,BenAmar13}.

The predictions of our model depart from those of the existing theories in several respects; here we mention two key differences. The first one is the detailed shape of the villi as the most elaborated nontrivial epithelial morpho\-logy. Our model villi are considerably taller and more tubular than those seen in the buckled elastic sheets~\cite{Shyer13,Ciarletta14,BenAmar13}, which are shorter and more similar to our previlli. As far as their height-to-width ratio is concerned, our villi are quite close to many real ones. Secondly, our model epithelium is devoid of an explicit support such as an elastic foundation. As such, it provides a basis for the understanding of villus-like formations characteristic of organoids as well as an insight into certain embryonic structures such as the archenteron in the sea urchin. Geometrically, these formations are clo\-sely reminiscent of the intestinal villi, yet they develop in absence of surrounding tissues, which implies that their morphogenesis can only be related to the intrinsic mechanics of the epithelium and to global constraints such as those fixing the lumen or blastocoel volume~\cite{Rozman20}. 

Of course, a generalized version of our model where the epithelium would rest on an elastic foundation could well be very interesting. We expect that such a substrate may define the wavelength of the folds~\cite{Hannezo11} as well as the length and the detailed shape of the villi, but it may also give rise to novel epithelial forms. In addition, the substrate would distinguish between the villi and the crypts. Our present model applies equivalently to both of these morphologies; the only difference between the apical and the basal surface is due to the differential tension, and thus a villus at a given $\beta-\alpha>0$ can also be regarded as a crypt at an equal and opposite $\beta-\alpha<0$. However, in an epithelium supported by an elastic substrate this equivalence no longer exists, with the villus growing out of the substrate and the crypt protruding into it.

Lastly, we emphasize that in our model, the existence of villi relies on the topological activity, the villi themselves having a nontrivial topological structure (Fig.~\ref{fig:1}h). This element of mechanics is absent in the continuum theories of epithelial form and it is plausible that the interplay of substrate elasticity and cell-level activity could result in novel morphologies if the former would interfere with the topology of the tissue.

\section{Conclusions}
\label{sec:5} 
The most important aspect of this study may well be the explicit comparison of the results of selected external and internal factors responsible for epithelial morphogenesis within a single-cell-population tissue. The strain arising from the competition between the tissue and any adjacent structure, be that it results from the growth of the epithelium attached to a substrate or from stress generated by the substrate, is a very plausible external source of buckling, yet we find that the buckled shape induced by it includes folds and bumps, but not true, well-developed villi. On the other hand, a high enough level of activity within the tissue and its ensuing fluidization is enough to promote villus and crypt formation provided that the apico-basal differential tension is large enough. Naturally, in a differentiated tissue consisting of biologically and mechanically distinct cell populations there may exist additional internal processes that contribute to the development of both fold and villus morphologies. 

The potential internal origin of villi and crypts also points to an intriguing hypothesis that certain other morphogenetic process in epithelial tissues may too be intrinsic rather than induced or driven by the environment. Branching, for example, relies on the formation of a Y~junction at the tip of the epithelial tube, and the saddle-shaped bases of the two growing buds are quite similar to those seen in the villi. From this perspective, budding is just a villus growing from the tip of an existing villus. Should this be possible, our vertex-model framework would provide the micromechanical interpretation of the elementary process of branching morphogenesis~\cite{Hannezo17}.
%
\bigskip

We thank P.~Mrak and M.~Rauzi for helpful discussions, and we acknowledge the financial support from the Slovenian Research Agency (research core funding No.~P1-0055 and projects No.~J2-9223 and Z1-1851).
\section*{Author contribution statement}
All authors were involved in the preparation of the manu\-script. All authors have read and approved the final manu\-script.

\section*{Data availability statement}
The datasets generated during and/or analyzed during the current study are available from the corresponding author on reasonable request.

\section*{Conflict of interest}
The authors declare that they have no conflict of interest.

\paragraph{Publisher’s Note} The EPJ Publishers remain neutral with
regard to jurisdictional claims in published maps and institutional
affiliations.
%
%

\end{document}


\maketitle
\thispagestyle{empty}
\setcounter{page}{0}
\newpage

\section*{Supplementary figures}
    	\begin{figure}[H]
		\centering
		\includegraphics{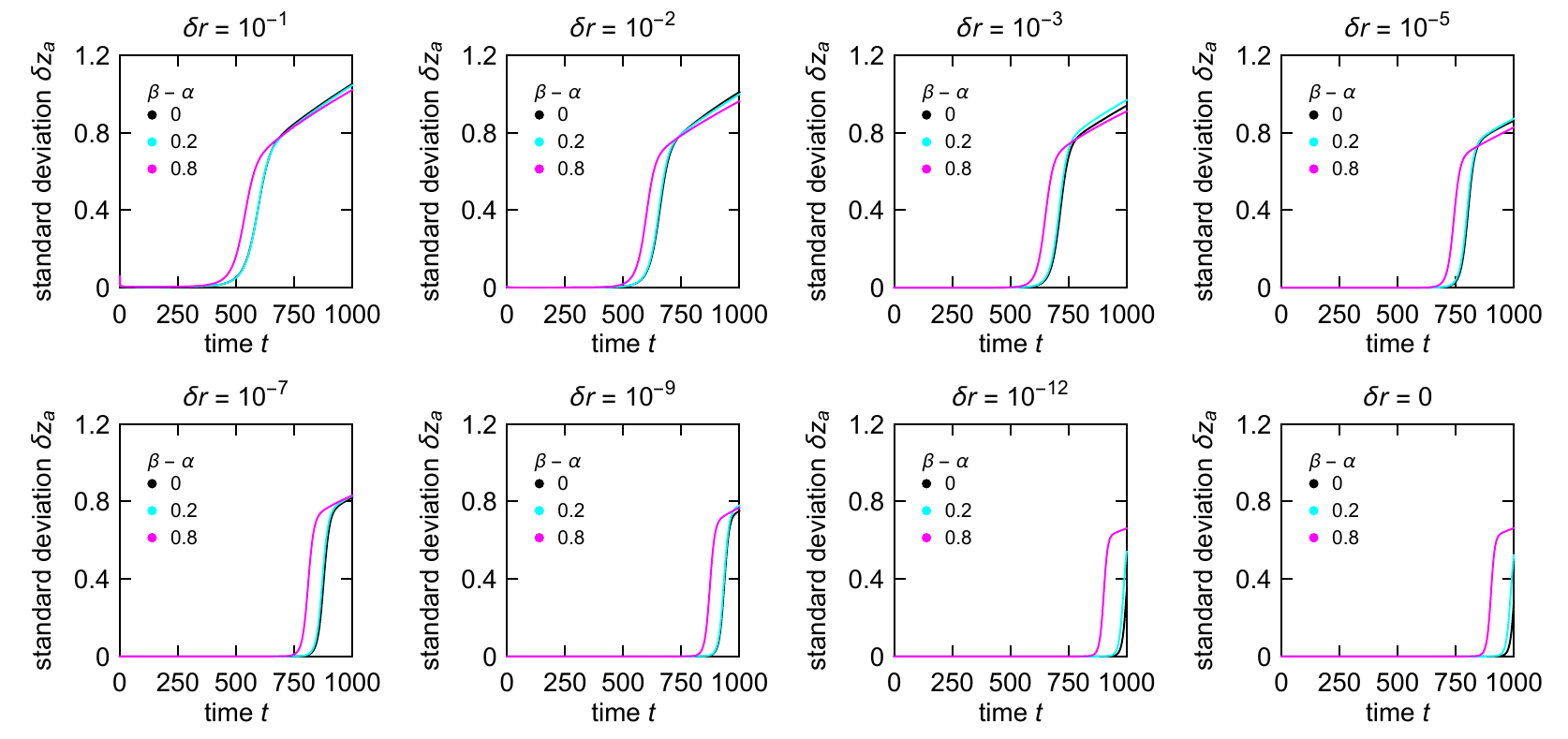}
		\caption{Standard deviation of the height of apical vertices $\delta z_a$ as a function of time at eight different values of the standard deviation of the initial perturbation of the vertices $\delta r$.}
	\end{figure}
	\begin{figure}[H]
		\centering
		\includegraphics{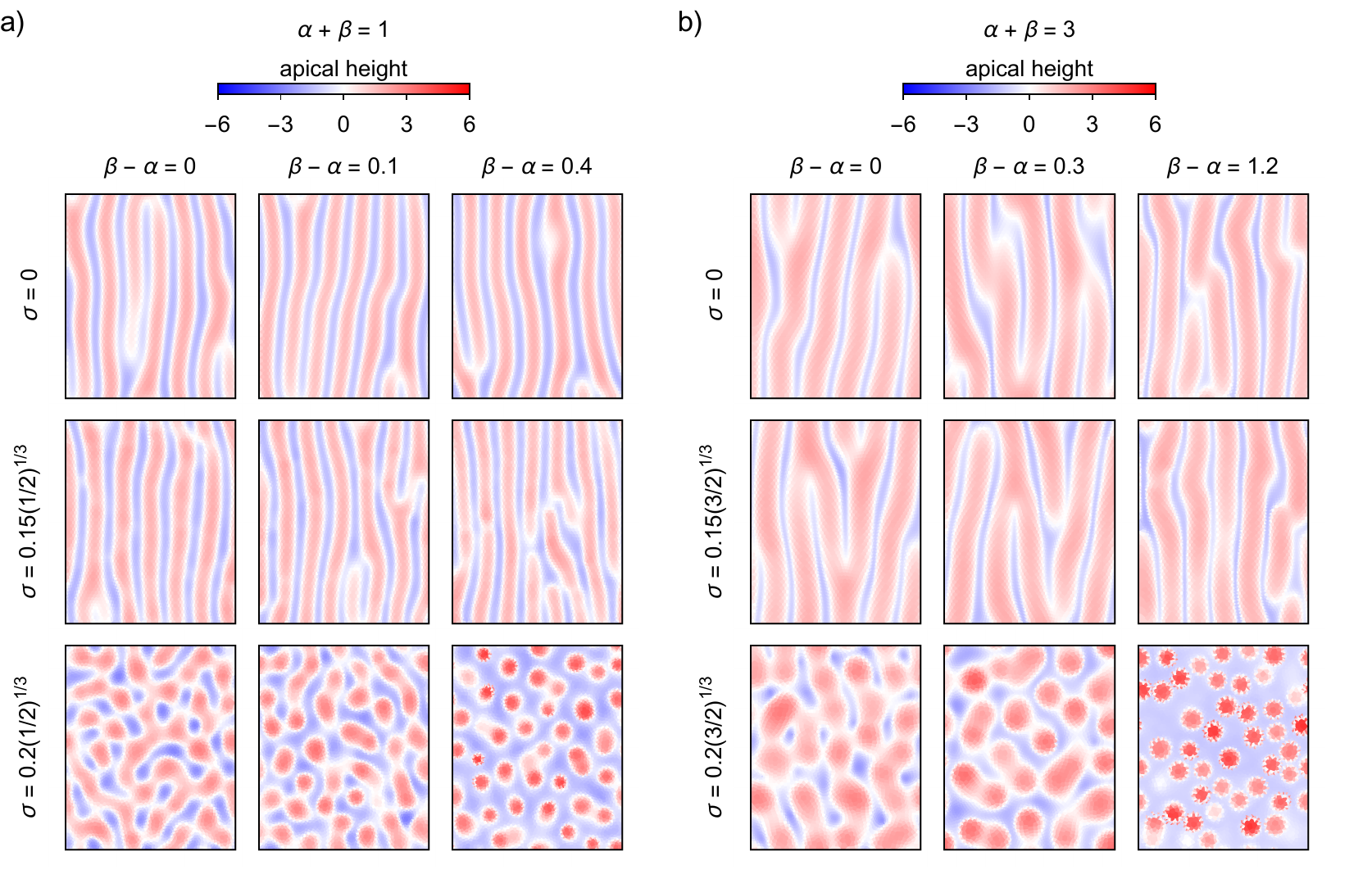}
		\caption{Height profiles of the apical surface of model epithelia formed by uniaxial compression in the horizontal direction by 27.75~\% for tissue tensions $\alpha + \beta = 1$~(a) and 3~(b). To facilitate comparison, we show the morphologies at differential tensions $\beta-\alpha$ such that the ratio $(\beta-\alpha)/(\alpha+\beta)$ in the left, middle, and right column in panels a~and~b as well as in Fig.~2a is 0, 0.1, and 0.4, respectively. We find that in the studied range of tensions, a reasonably good way to select equivalent tension fluctuations for comparison at different $\alpha + \beta$ is to rescale the value of $\sigma$ by the characteristic in-plane dimension and the effective mean line tension.   The characteristic in-plane dimension is given by the linear size of the equilibrium apical or basal side of a hexagonal-prismatic cell equal to $a_0 = \left(\sqrt{3}/2\right)^{1/3}(\alpha+\beta)^{-2/3}$. To determine the effective mean line tension, we follow Ref.~\cite{Krajnc18}. In a flat tissue with a constant cell height $h_0$ and consequently a constant cell apical and basal area $a_0$, we recast the non-active part of the energy [Eq.~5 of the main text] as $w = \sum_{i=1}^{N_c} \left[\alpha a_0+\beta a_0+h_0p^{(i)}/2\right]={\rm const.} + \sum_{j=1}^{N_e} h_0 l^{(j)}$, where $p^{(i)}$ is the perimeter of the $i$-th cell and $l^{(j)}$ is the length of the $j$-th junction. From this expression we can see that the cell height $h_0$ plays the role of the mean effective line tension; in turn, $h_0$ can be estimated by the equilibrium value in a hexagonal-prismatic cell $h_0 = \left(\sqrt{3}/2\right)^{-1/3}(\alpha+\beta)^{2/3}$. Therefore, we chose $\sigma$ values for comparison by the relation $\sigma(\alpha+\beta)/(\alpha+\beta)^{1/3}\approx\sigma(\alpha'+\beta')/(\alpha'+\beta')^{1/3}.$ 
The morphologies obtained at the appropriately rescaled differential tensions and tension fluctuations for $\alpha+\beta=1$, 2, and 3 are very similar. Note that while the profiles shown here are plotted in same-size rectangles so as to facilitate comparison, the actual area of the more columnar tissue with $\alpha+\beta=3$ is smaller  than in the tissue in Fig.~2a, whereas the area of the more cuboidal tissue with $\alpha+\beta=1$ is larger than in that in Fig.~2a.}
	\end{figure}
	\begin{figure}[H]
		\centering
		\includegraphics{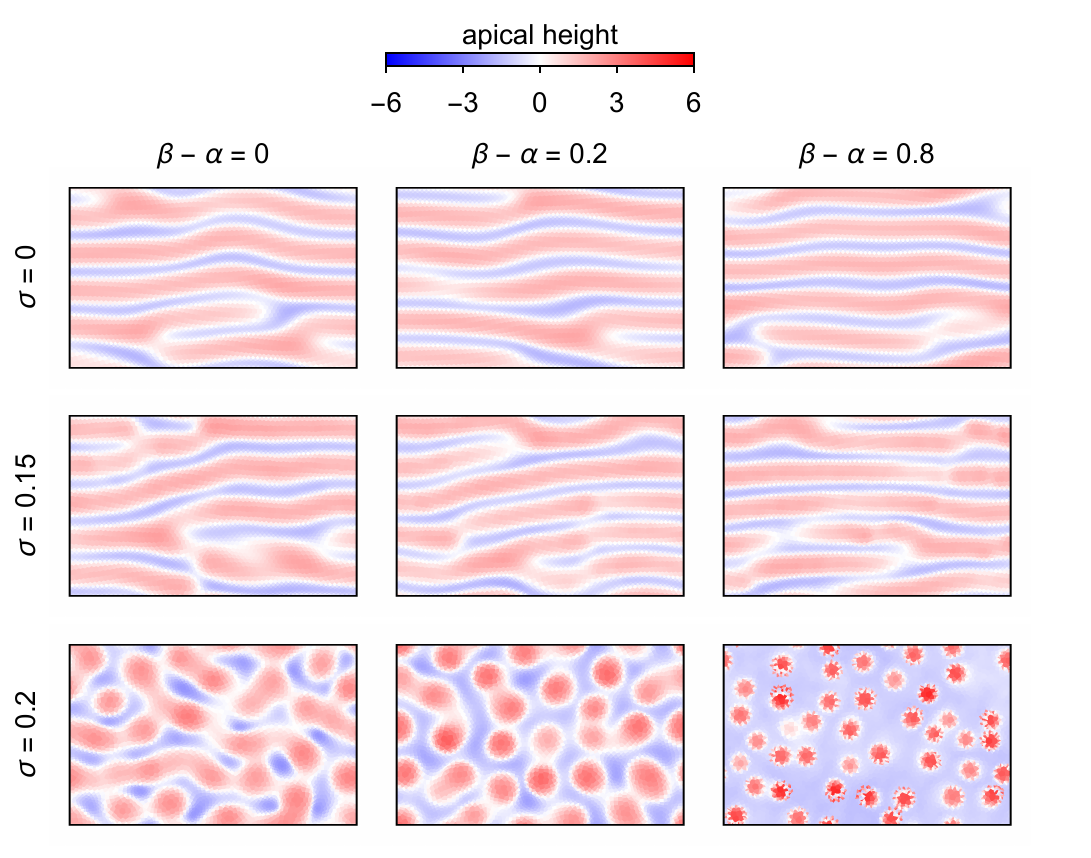}
		\caption{Height profiles of the apical surface of epithelial morphologies at $\alpha + \beta = 2$ formed by uniaxial compression along the $y$ axis by 27.75~\%. The morphologies obtained are equivalent to those that appear under compression along the $x$ axis.}
	\end{figure}

	\begin{figure}[H]
		\centering
		\includegraphics{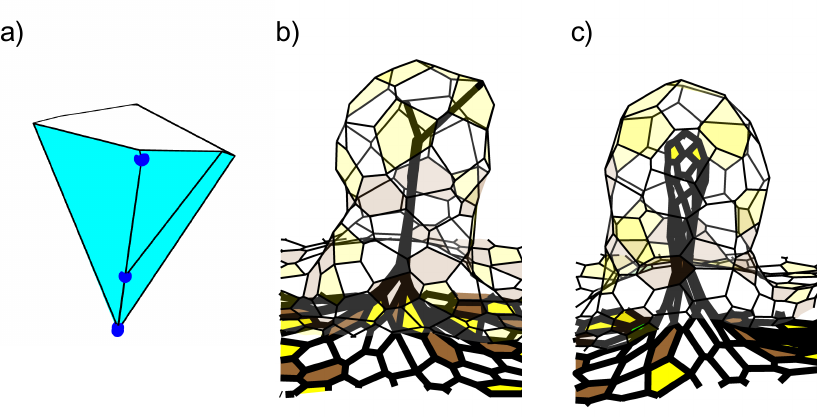}
		\caption{Example of a pyramidal cell at the tip of the villus, with vanishingly small basal side and several basal vertices shifted towards the apical side~(a). The apical side is shown in white, the lateral sides are cyan, and the basal vertices are indicated by blue points. Panel~b shows an example of a villus with spikes at the basal side of cells at the tip, whereas that in panel~c is free of such spikes as it was computed using a generalized version of our vertex model which includes an auxiliary term that prevents the appearance of the spikes by imposing an energy penalty on small basal areas. For simplicity, the villi presented in panels~b and c were computed in a patch of $32\times32=1024$~cells with $\alpha+\beta = 2$, $\beta - \alpha = 0.8$, and $\sigma = 0.2$; $\varepsilon_x = 27.75~\%$ and $\varepsilon_y=0$.}
	\end{figure}
	\begin{figure}[H]
		\centering
		\includegraphics{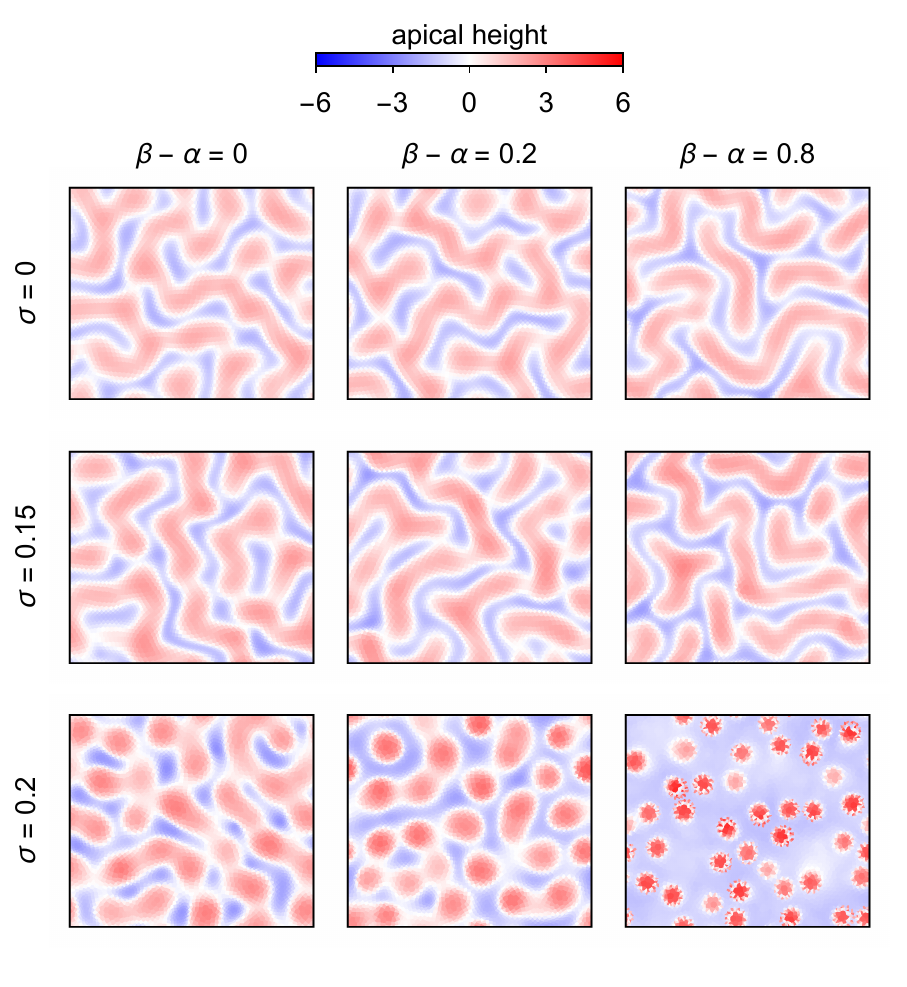}
		\caption{Height profiles of the apical surface of epithelial morphologies at $\alpha + \beta = 2$ formed by isotropic compression by $\varepsilon_x = \varepsilon_y=15~\%$.}
	\end{figure}
    \begin{figure}[H]
		\centering
		\includegraphics{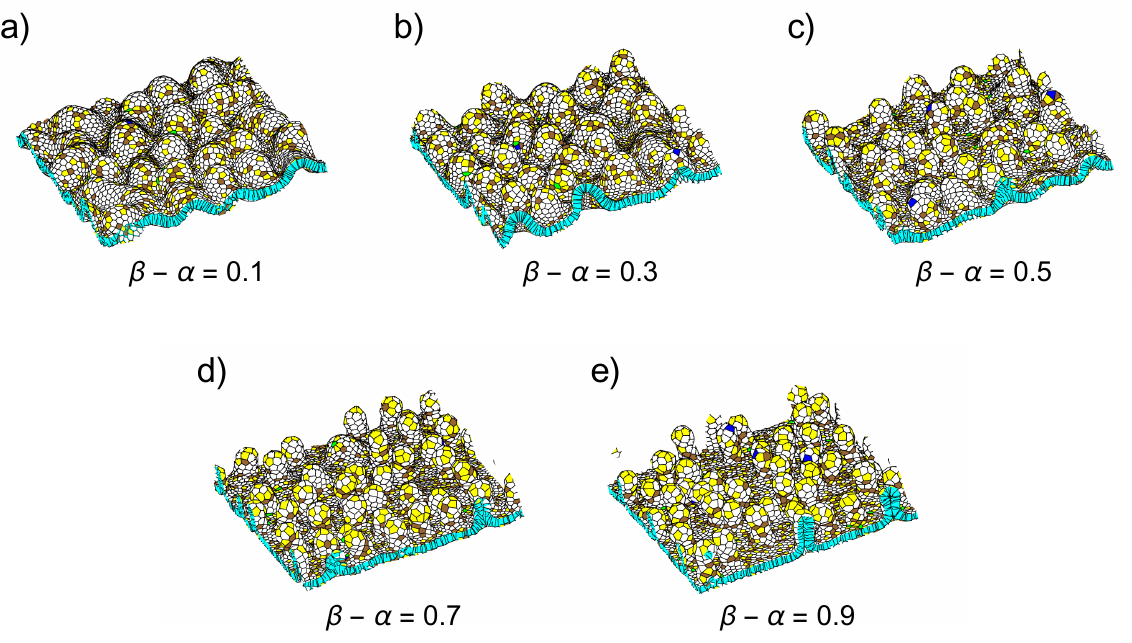}
		\caption{Apical surface view of computed epithelial morphologies following uniaxial compression with $\alpha+\beta=2$, $\sigma =0.2$, $\varepsilon_x = 27.75~\%$, and $\varepsilon_y = 0$, at differential tension $\beta-\alpha = 0.1$ (a), $\beta-\alpha = 0.3$ (b), $\beta-\alpha = 0.5$ (c), $\beta-\alpha = 0.7$~(d), and $\beta-\alpha = 0.9$ (e).}
	\end{figure}
	\newpage

\section*{Supplementary movies}
\noindent\textbf{Movie~S1.} Temporal evolution of longitudinal folds with $\alpha+\beta=2$, $\beta-\alpha=0$, and $\sigma=0$ under uniaxial compression with $\varepsilon_x = 27.75~\%$ and $\varepsilon_y = 0$.\\

\noindent\textbf{Movie~S2.} Temporal evolution of previlli with $\alpha+\beta=2$, $\beta-\alpha=0$, and $\sigma=0.2$ under uniaxial compression with $\varepsilon_x = 27.75~\%$ and $\varepsilon_y = 0$.\\

\noindent\textbf{Movie~S3.} Temporal evolution of villi with $\alpha+\beta=2$, $\beta-\alpha=0.8$, and $\sigma=0.2$ under uniaxial compression with $\varepsilon_x = 27.75~\%$ and $\varepsilon_y = 0$.

\noindent\textbf{Movie~S4.} Temporal evolution of labyrinthine folds with $\alpha+\beta=2$, $\beta-\alpha=0$, and $\sigma=0$ under isotropic compression with $\varepsilon_x = \varepsilon_y = 15~\%$.\\